\documentclass[aps,prx,twocolumn,showpacs,superscriptaddress,groupedaddress]{revtex4-2}

\usepackage[caption=false]{subfig}
\usepackage{graphicx} 
\usepackage{mathrsfs}
\usepackage{amsfonts}
\usepackage{amsmath}
\usepackage[noabbrev]{cleveref}
\usepackage[toc,page]{appendix}
\usepackage{import}
\usepackage[nice]{units}
\usepackage{float}
\usepackage{color}

  \def\a{\alpha} \def\b{\beta}
\def\g{\eta} \def\d{\delta}

\newcommand\mymat[1]{\begin{pmatrix} #1 \end{pmatrix}}

\begin{document}
\title{Cumulative geometric frustration in physical assemblies}
\date{\today}
\author{Snir Meiri}
\author{Efi Efrati} 
\email{efi.efrati@weizmann.ac.il} 
\affiliation{Department of
Physics of Complex Systems, Weizmann Institute of Science, Rehovot
76100, Israel}

\begin{abstract} 
Geometric frustration arises whenever the constituents of a physical assembly locally favor an arrangement that cannot be realized globally. Recently, such frustrated assemblies were shown to exhibit filamentation, size limitation, large morphological variations and other exotic response properties. While these unique characteristics can be shown to be a direct outcome of the geometric frustration, some geometrically frustrated systems do not exhibit any of the above phenomena.
In this work we exploit the intrinsic approach to provide a framework for directly addressing the frustration in physical assemblies. The framework highlights the role of the compatibility conditions associated with the intrinsic fields describing the physical assembly. We show that the structure of the compatibility conditions determines the behavior of small assemblies, and in particular predicts their super-extensive energy growth exponent. We illustrate the use of this framework to several well known frustrated assemblies.
\end{abstract}
\maketitle

\section{\label{intro}Introduction\protect\\}
The ground state of an assembly of identical particles endowed with short range interactions is expected to reflect the symmetries of space and of its constituents. Hard discs with short range attraction in the plane will pack tightly in a six-fold symmetric order, such that the centers of the discs will form the vertices of a unilateral triangular lattice. This uniform order, however, cannot persist if the discs are packed in curved space, e.g. the surface of a much larger sphere. In such a curved space the sum of the internal angles in each of the formed triangles deviates from the preferred value of $\pi$ by an angle deficit that scales as $\Delta\sim\tfrac{d^{2}}{\rho^{2}}$, where $d$ is the inter-particle distance and $\rho$ is the radius of the larger sphere. While this angle deficit is identical in all the triangles, it leads to spatial gradients in the packing fraction of the bulk ground state \cite{Gra16}. These inevitable strain gradients are associated with a super extensive elastic energy contribution, in which the elastic energy per particle grows as the area of the domain increases \cite{ESK13}, and favors the formation of narrow filamentous domains over isotropic bulks \cite{MPNM14,Gra16,ESK13}.

The above phenomenology of frustrated assemblies was verified numerically and analytically for defect free crystals growing in a uniformly curved geometry \cite{SG05}, as well as observed experimentally in a system of colloids confined to a spherical interface and endowed with very short range attraction \cite{MPNM14}. The tendency to form filaments, the super extensive elastic energy and the non-trivial dependence of the domain size on the line tension can all be attributed to the mismatch between the attempted (vanishing) Gaussian curvature of the material manifold, and the non-zero Gaussian curvature of the ambient space in which it is embedded. This provides a natural geometric charge that accumulates and serves as a source term for the spatial strain gradients in the material. 
Similar behaviors appear in many other systems including filament bundles \cite{BG12,HG17}, liquid crystals \cite{Set83,NE18}, chiral stiff rod-like colloids \cite{GBZ+12} and twisted molecular crystals \cite{HAS+18,LSV+20}. For some of these systems, the geometric charge associated with the frustration has not been identified. Recent works aim to provide a unified framework to describe geometric frustration in these diverse systems \cite{Gra16,HG20}.  

There are, however, geometrically frustrated systems, such as the Ising anti-ferromagnet on a triangular lattice \cite{Wan50}, that do not exhibit any of the above traits. The lowest energy per nearest neighbor edge cannot be realized simultaneously on all three edges of a single triangular facet, giving rise to frustration. However, the lowest energy compromise on a single facet can be realized uniformly throughout the lattice, giving rise to a trivial extensive energy scaling.

Gluing the faces of two oppositely curved thin elastic cylindrical sheets gives rise to a similar type of frustration; the two thin sheets cannot simultaneously realize their desired curvature, yet the ground state compromised assumed by flattening their curvatures is uniform, as depicted in figure 1.  

\begin{figure*}[]
\includegraphics[width=17.8cm]{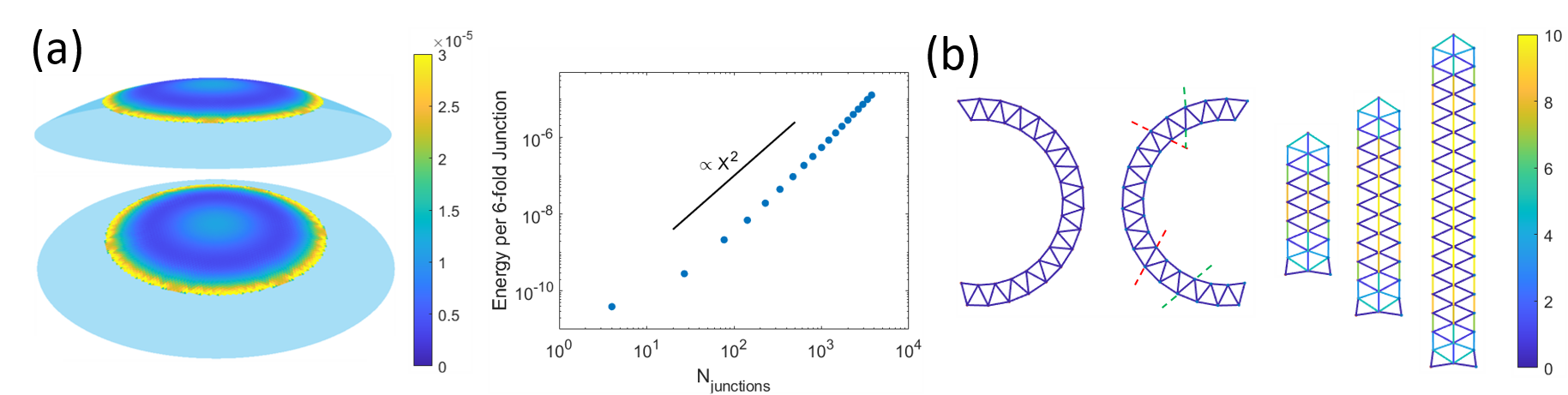}
\caption{ (a) Numeric simulation of embedding of an elastic triangular lattice on a sphere. The Radius of the sphere is 10 (a.u) and the preferred distance between neighboring vertices is 0.1 (a.u). (Left) Top view of an isotropic domain with color of each triangle representing its elastic energy according to the colorbar (similar figure was shown in \cite{HBBG16}). (Right) Logarithmic plot of the energy per 6-fold junction in the lattice vs. the number of such junctions in the domain. Scaling line of $X^{2}$ is added to guide the eye. (b) Gluing two bent triangulated sheets. Simulation results for non-glued sheets and 3 different lengths of sheets glued with 8, 12 and 18 vertices per sheet's face (left to right). The range of the cut sheets is marked by red and green dashed lines. The color of each edge represents its elastic energy according to the colorbar to the right. }
\label{fig:intro}
\end{figure*}


The crystalline packing of attractive discs on curved surfaces, the Ising anti-ferromagnet on triangular lattice and the glued cylindrical thin sheets all display geometric frustration: Geometric constraints prevent the locally preferred ground state from being realized uniformly throughout the system. However, the resulting phenomena differ substantially between the two cases suggesting that the former one and latter two belong to different classes of geometrically frustrated systems. While the first system features super extensive behaviour which drives the assembly towards fillamentation and endows it with exotic properties, the latter two systems feature an extensive ground state energy. What then distinguishes between the different types of frustration?

In the above examples it was necessary to solve the systems' ground state to distinguish if the frustration was cumulative, causing the increased buildup of strains as the system grew in size, or non-cumulative leading to a uniform solution with extensive energy. Finding this ground state is further complicated by the lack of a stress-free state in frustrated systems.

In this work we adopt the intrinsic approach which overcomes the lack of a stress free state in frustrated systems, and clarifies the origin of the associated frustration. In this approach strains are not a measure of the deviation from a stress free configuration, but rather measure the deviations from the mutually contradicting local tendencies of the system. The inability of the system to simultaneously comply with all the locally desired tendencies is captured by the compatibility conditions any realizable configuration must satisfy.

These compatibility conditions in turn, contain the information on the possible low-energy resolutions of frustration in the system for small domains. Their form allows us to distinguish if the frustration in a given system is cumulative or non-cumulative, and furthermore provides a measure of the strength of the frustration in the form of the super extensive energy scaling exponent without explicitly solving for the ground state. We use this framework to study the frustration in several specific well known frustrated systems.

\section{\label{systems}
Determining frustration strength and
Distinguishing cumulative from non-cumulative frustration\protect\\}
Super-extensive energy is not-uncommon for systems with long range interactions, where every particle interacts with all other particles. For example, a uniform electric charge density $\rho$ in a spherical domain of radius $r$ is associated with the energy $E\propto \rho^2 r^5\propto M^{5/3}$, where $M\propto r^3$ is the total mass of the charges \cite{Fey74}. For short range interactions, one expects that every constituent will affect only those in its immediate vicinity. However, the cooperative nature of the ground state in systems displaying cumulative geometric frustration causes these systems to display long range behavior and exotic morphological response properties \cite{LW17,Gra19,AAMS14,ZGDS19,LS16,GSK18}.

We adopt the super-extensive energy as the defining characteristic of cumulative frustration. Systems in which the local resolution of the frustration can be propagated globally to the entire system will be associated with non-cumulative frustration, and their ground state energy will show extensive scaling, $E\propto M$.  
Systems with short range interactions in which the local optimal compromise at some small region of the system cannot be repeated in its vicinity necessitating a more energetically expensive resolution of the frustration will show super-extensive energy scaling, and be associated with cumulative geometric frustration. The strength of the cumulative frustration is captured by the super-extensive ground state energy exponent
\[
\lambda=\left.\frac{\partial \log{E}}{\partial \log{M}}\right|_{M\to0}.
\]

To distinguish between cumulative and non-cumulative frustration in continuous systems we examine their Hamiltonian and its ability to support ground state solutions with spatially uniform energetic cost. In practice this requires finding new variables that \textit{(i)} fully characterize the state of the system, and \textit{(ii)} allow to express the Hamiltonian in a local form, i.e. containing no spatial derivatives. These new variables are, however, not independent of each other as they are derived from common native variables and their gradients. This interdependence manifests in functional constraints the new variables must satisfy which are termed the compatibility conditions. 
For any frustrated system the compatibility conditions preclude the simultaneous minimization of all the terms in the Hamiltonian.
The structure and nature of the compatibility conditions determine the class of frustration. Local (non-differential) compatibility conditions give rise to non-cumulative frustration whereas if all the compatibility conditions contain differential relations they are indicative of cumulative frustration; We next make this statement more precise, and in particular deduce $\lambda$ directly from the compatibility conditions.  

For clarity, in what follows we first derive the super-extensive energy exponent for systems that are small enough, whose energy landscape is non-stiff, and that are spatially isotropic. The notions of small enough isotropic domains, and energy stiffness are then explained, followed by discussing the process of elimination of stiff energy directions for the relevant systems, to allow applying the same analysis.

\subsection{Calculating the super-extensive energy exponent}
Consider a Hamiltonian that depends on $n$ fields and their derivatives:
\[
\mathcal{H}\left(\phi_1,\phi_2,...,\phi_n, \nabla \phi_1,...,\nabla \phi_n,\nabla\nabla \phi_1,... \right)
\]
We associate new fields $\psi_i$ with each of the fields and fields' gradients that appear in the Hamiltonian. The variables $\psi_i$ bring the Hamiltonian to a local form (containing no derivatives), yet must comply with $k$ distinct differential constraints that relate them and their derivatives to each other:
\[
G_i\left(\psi_1,\psi_2,...,\psi_n, \nabla \psi_1,...,\nabla \psi_n,\nabla\nabla \psi_1,... \right)=0
\]
We group the fields to a vector $\Psi$ and identify the local ground state of the Hamiltonian with the values $\bar\Psi$. These preferred values could vary in space or assume constant values depending on the context.
We expand the Hamiltonian to second order in the generalized strain, $\varepsilon=\Psi-\bar{\Psi}$, which measures the deviations of these fields from their locally preferred values. We then express the compatibility conditions in terms of this generalized strain by substituting $\Psi=\varepsilon+\bar{\Psi}$. For systems displaying cumulative frustration the compatibility conditions will take the form of partial differential relations. In general, for all frustrated systems the compatibility conditions preclude the state $\varepsilon=0$ from being globally achieved. Thus, expanding the strain in orders of the spatial coordinates, e.g. for two dimensions $\varepsilon\approx \varepsilon_{0,0}+\varepsilon_{1,0}x+\varepsilon_{0,1}y+...$, the compatibility conditions preclude setting all the coefficients $\varepsilon_{i,j}$ to zero. For small domains the lowest order terms dominate the rate of growth of the strain, and to minimize the energy we seek to eliminate as many low order coefficients as the compatibility conditions allow. If already $\varepsilon_{0,0}$ cannot be set to zero, then the system will exhibit an optimal compromise with extensive energy scaling. If this coefficient can be set to zero, then the first coefficient that cannot be set to zero $\varepsilon_{i,j}$ where $i+j=\eta$ will lead to a strain that scales as
\[
\varepsilon\propto r^\eta,
\]
for small enough isotropic domains that are smaller than the geometric length-scale arising from the compatibility conditions (a similar definition of the exponent $\eta$ may be found in \cite{HG20}). 
The associated energy in this case will scale as 
\[
E\propto \int d^dr \epsilon^2\propto r^d r^{2\eta}\sim M^{1+\frac{2\eta}{d}}
\]
For example if the compatibility conditions amount to a single differential equation of order $p$, and the locally preferred values of the fields $\bar{\Psi}$ are smooth in the domain considered, then $p\le\eta$. To determine the exact value of $\eta$ we expand the compatibility condition in terms of the spatial coefficients of the strain. If the zero-order equation is non-homogeneous in $\epsilon_{i,j}$ then $\eta=p$. If it is homogeneous, then the first non-homogeneous order of the compatibility condition, $q$,  will yield $\eta=p+q$. In general when there are multiple compatibility conditions the lowest term in the vector $\epsilon_{i,j}$ that cannot be set to vanish will determine the exponent $\eta$. The reader is referred to appendices \ref{app:Elas2D} and \ref{app:Elas3D} for examples. 

In general, the structure and differential order of the compatibility
conditions determine the rate of growth of the associated energy of the optimal compromise of a frustrated system. Note, that systems that display non-cumulative frustration are associated with local compatibility conditions that are characterized by $\eta=0$, which indeed by the formula above yield $E\propto M$.

\subsection{The notion of sufficiently small domains}
The exponent $\lambda$ was formally defined in the limit $M\to 0$, and in practice remains valid provided the domain is sufficiently small.  The restriction to small domains is required in order to identify the spatial growth rate of the strain with the power law growth $r^\eta$. $\eta$ in turn is determined by the first order in the spatial expansion of $\epsilon$ that cannot be set to vanish due to a non-homogeneous term in the compatibility condition. Determining the exact regime of validity requires comparing the locally preferred value of $\Psi$ with the magnitude of its gradient prescribed by the compatibility conditions. In many systems the locally preferred values $\bar{\Psi}$ and the corresponding compatibility conditions contain a single or small number of intrinsic length-scales associated with the frustration. Thus the restriction for small domains (which is geometric in nature) reduces to requiring that diameter of the isotropic domain is smaller than the geometric length-scales associated with the frustration.

\subsection{Stiff Hamiltonians and anisotropic domains}
The Hamiltonian expanded to second order in the local strains reads
\[
\mathcal{H}=\int_\Omega (\Psi-\bar{\Psi})^T \chi (\Psi-\bar{\Psi})
dV=\int_\Omega \varepsilon^T \chi \varepsilon
dV.
\]
In the above energy growth rate considerations all coordinates, fields and field gradients were treated equally. For stiff systems, in which the eigenvalues of the constitutive coefficient matrix $\chi$ are not comparable, the stiff (highly energetically penalized) distortion fields must be treated differently. Similarly, if the domain $\Omega$ is highly anisotropic, gradients along the small direction will be associated with reduced energetic cost, leading again to a stiff energetic landscape. In both cases the previous analysis  
breaks.  
 
The partial freezing of degrees of freedom that results from highly anisotropic domains or the existence of stiff directions in the Hamiltonian may effectively change the differential order of the compatibility conditions and therefore of the scaling behavior they predict.
In order to apply the analysis presented above to such stiff and anisotropic systems we first take the limit in which
the energetic penalty in the stiff direction diverges, or alternatively the limit in which the small dimension vanishes. This leads to a limiting Hamiltonian that is dimensionally reduced or relates fewer fields. If this reduced Hamiltonian no longer contains stiff directions, and the dimensionally reduced domain is isotropic, we may calculate the super extensive exponent $\lambda$ as described above. 
The resulting compatibility conditions, and energy-scaling exponents predict the behavior of the Hamiltonian in the stiff regime and for highly anisotropic small enough domains. We note that the length-scales used to determine if the domain is small enough are those prescribed by the reduced Hamiltonian and reduced compatibility conditions. 
For example in section \ref{bi-rib} we analyze thin elastic sheets of moderate thickness using the full Hamiltonian, while for very thin sheets, the reduced Hamiltonian obtained in the vanishing thickness limit predicts a different energy scaling. 

While the limiting behavior of such Hamiltonians may be straightforward to obtain, the analysis of the possible flow in the abstract space of compatibility conditions remains outside the scope of the present work.



\subsection{Frustration saturation}
The individual building blocks of a frustrated assembly are assumed to be relaxed prior to assembly, it is only their rearrangement upon assembly that gives rise to frustration. The energy per particle at the onset of frustration, $e_{min}=E_{min}/n$, where $n$ is the number of particles, is the smallest energy scale associated with the frustration. For systems displaying cumulative frustration the energy associated with the subsequent addition of building block is higher and increases with size. This increase, however cannot persist indefinitely, as it will give rise to an arbitrarily large energy per particle.
Most physical assemblies will be associated with a finite local frustration resolution energy: One could simultaneously satisfy all the compatibility conditions by choosing some constant state $\Psi^*$ that is far from $\bar\Psi$. This will result in an extensive energetic cost, with energy per particle $e^*$ that is typically very high compared with $e_{min}$. There are cases, however, in which $e^*\sim e_{min}$. In these cases, the frustration saturates at the level of individual building blocks, and will not lead to a super-extensive energy, nor play an important role in shaping the assembly. This is the case of the Ising anti-ferromagnet, as well as a variety of similar frustrated spin systems \cite{RL19}. For continuous systems that display cumulative frustration, one cannot associate a minimal frustration energy with the assembly, and sufficiently small domains will always display a super-extensive behavior. 
\\
To understand the outcome of the frustrated assembly of many building blocks and to predict if the resulting assembly will be self-limiting, become defect ridden or develop into a uniformly frustrated bulk, thermodynamic considerations must be taken into account. Concerned with equilibrium mechanisms of self-limiting assembly \cite{HG20},
Hagan and Grason proposed a new measure for predicting the thermodynamic outcome of such assemblies, termed the accumulant:
\[
\mathcal{A}\left(W\right)=W[\varepsilon_{\infty} -\varepsilon_{ex}\left(W\right)]-\Sigma,
\]
where $\varepsilon_{\infty}$ is the energy per unit volume in the infinite bulk, $\varepsilon_{ex}\left(W\right)$ is the excess energy per unit volume associated with the frustration, $W$ is the length in the potentially self-limiting dimension, and $\Sigma$ is the surface energy density. The regime in which the accumulant $\mathcal{A}$ increases with $W$ can support a finite equilibrium self-limiting assembly. If, however, $\mathcal{A}$ diminishes with $W$ then in this regime the system cannot support a size limited finite assembly and will form a bulk. This behavior is dominated by the way the frustration energy approaches saturation. This behavior is not predicted with the tools provided here, aimed at characterizing only the frustration energy at the onset of assembly.

\section{Continuous frustrated assemblies}
In this section we implement the intrinsic approach and the framework presented above to quantitatively examine the frustration in four well studied continuous systems that exhibit geometric frustration. These consist of uniformly and isotropically frustrated elastic structures in two and three dimensions, frustrated liquid crystals in two dimensions and the bi-stable elastic bi-layer which could be considered in several distinct limits, leading to distinct energy exponents.

\subsection{Uniform and isotropic frustration in elastic solids in two and three dimensions}
Consider the embedding of an elastic disc with the geometry of $\mathbb{S}^2$ in the plane. One may think of this problem as flattening an infinitely thin spherical cap between two flat glass plates and examining the in-plane stresses. In the standard elastic description the energy is quadratic in the strain measured using gradients of the displacement vector. However, in the present case no stress free configuration exists in the plane, and a displacement vector cannot be defined. The framework of metric elasticity, is particularly suited for such residually-stressed systems, as it overcomes this difficulty by measuring strains with respect to a reference metric. Moreover, it brings the elastic energy to the desired local form:
\begin{equation}
    E=\int\int \mathcal{A}^{\a\b\g\d}(a_{\a\b}-\bar{a}_{\a\b})(a_{\g\d}-\bar{a}_{\g\d}) d\bar{A},  
    \label{eq:elasticEnergyS2}
\end{equation}
where Greek indices assume the values $\{1,2\}$, $\bar{a}_{11}=1,\,\bar{a}_{12}=\bar{a}_{21}=0$ and 
$\bar{a}_{22}=\sin(r)^2$ is the locally preferred reference metric, and the elasticity tensor $\mathcal{A}$ and the area element $d\bar{A}$ depend only on the reference metric and material properties and not on the configuration assumed \cite{ESK09}. The two dimensional metric, $a$, fully describes the configuration of the system, yet only metrics of vanishing Riemannian curvature can describe the sought planar solution. The latter constraint leads to a compatibility condition in the forms of a single non-linear second order partial differential equation in the components of $a$, see appendix \ref{app:Elas2D}. Consequently, $\eta=2$ and the energy associated with embedding an isotropic domain grows as $E\propto M^3$. Similar results are obtained for embedding a flat surface in uniformly curved space such as growing a defect free crystal on the surface of a sphere \cite{MPNM14,SG05,BG12}, as depicted in Figure \ref{fig:intro}.

The elastic strains associated with the geometric frustration in the above system need not be very large. We thus expect alternative elastic descriptions to yield the same results to leading order. However, these may have significantly different structures. For example, the deformation gradient approach to residually stressed bodies \cite{GN71,Hog97}, employs the deformation gradient $F=\nabla \mathbf{r}$ as the basic variable describing the configuration of the system. The resulting compatibility condition, $\nabla\times F=0$ is of first order. However, careful examination reveals that the first order in the compatibility condition is trivial and the first non-homogeneous term remains of second order, predicting again $\eta=2$. 
See appendix \ref{app:Elas2D}.

\begin{figure}[h]
\includegraphics[width=6.5cm]{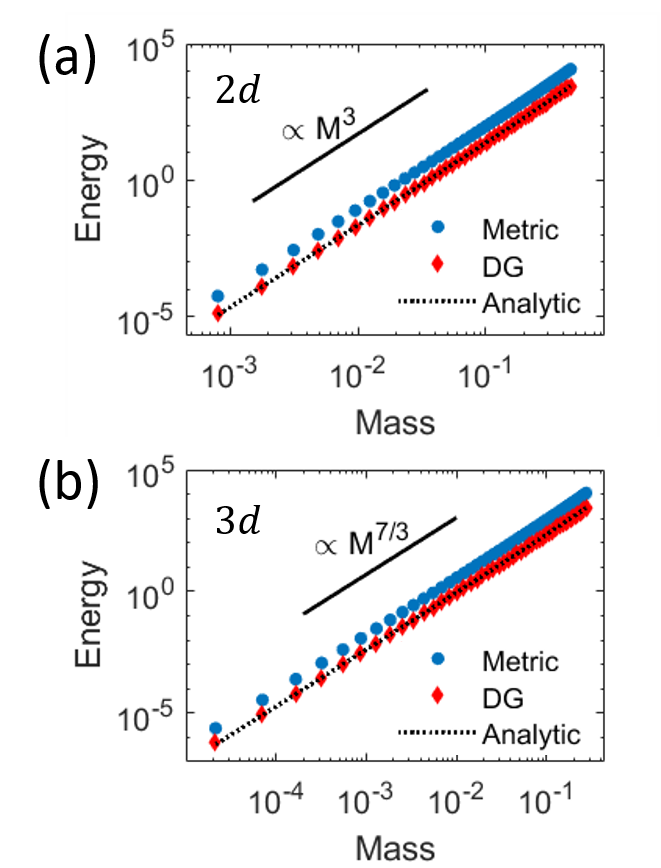}
\caption{ Numeric results and analytic solutions of embedding isotropic geodesic domains with positive Gaussian curvatures in Euclidean space. (a) Results of the embedding a geodesic disc from $\mathbb{S}^{2}$ in $\mathbb{E}^{2}$. (b) Results of the embedding a geodesic sphere from $\mathbb{S}^{3}$ in $\mathbb{E}^{3}$. Numeric results for the minimization of the energy functional resulting from the metric description are marked in blue filled circles while the results attained in the case of the deformation gradient (DG) description are marked in red diamonds. Analytic solution is marked in dashed line. Notice the logarithmic scale in both mass and energy. Scaling lines of $M^{3}$ and $M^{7/3}$ are added to guide the eye.}
\label{fig:elastic}
\end{figure}

Considering embedding a piece of the uniform and maximally symmetric positively curved manifold $\mathbb{S}^3$ in Euclidean space $\mathbb{E}^3$ leads to an elastic energy of the form of \eqref{eq:elasticEnergyS2}, with $a$, $\bar{a}$, $\mathcal{A}$ and $d\bar{A}$ replaced by their three dimensional variants. In three dimensions the elastic compatibility conditions consist of three non-linear second order differential equations in the components of the metric.  A straightforward calculation shows that this too yields $\eta=2$ which in turn results in $E\propto M^{7/3}$. Similarly to the two dimensional case, the deformation gradient approach gives rise to a system of first order compatibility conditions, yet upon explicit substitution predicts the same exponent. Figure \ref{fig:elastic} shows the comparison between the elastic energy computed numerically for both metric elasticity and the deformation gradient approach and the exponent predicted from the structure of the compatibility conditions. The general result $E\propto M^{1+4/d}$ was obtained through scaling arguments and a formal asymptotic expansion in \cite{AKM+16}, and rigorously proven through $\Gamma$-limits in \cite{MS19}.
For more details see appendix \ref{app:Elas3D}.

\subsection{Frustrated liquid crystals}
Next, we consider the following general Hamiltonian of two fields in the plane:
\begin{equation} \label{eq:genHam2Var}
H=\int[K_1(\Psi_1-\overline{\Psi}_1)^2+K_2(\Psi_2-\overline{\Psi}_2)^2]dA,
\end{equation}
where $K_1$ and $K_2$ are constant coefficients. Without knowledge of the form of the compatibility conditions for $\Psi$, the ground-state energy-scaling cannot be addressed.

Two dimensional liquid crystals are characterized by a local unit director field 
\(\hat{n}=(\cos (\theta),\sin(\theta))\).
For liquid crystals we may identify \eqref{eq:genHam2Var} with the Frank free energy where 
$\Psi_1=\hat{n}\wedge\nabla\theta=\nabla\cdot \hat{n}=s$ is the local splay, and
$\Psi_2=|\hat{n}\cdot\nabla\theta|=b$ is the bend of the director field. The compatibility condition in this case reads  \cite{NE18}:\\
\begin{equation} \label{eq:CompBCLC}
b^2+s^2+\hat{n}\cdot{\nabla}s-{\hat{n}}_\perp\cdot\nabla{b}=0.
\end{equation}
Bent-core liquid crystals confined to the plane are known to display geometric frustration \cite{Mey76}, and are associated with the reference values $\bar{\Psi}=\{\bar{s},\bar{b}\}=\{0,b_0\}$.
With respect to these values
the first order compatibility condition \eqref{eq:CompBCLC} is non-homogeneous in the strain (which in turn necessitate non-vanishing gradients) yielding $\eta=1$. Correspondingly the energy of the optimal compromise for small domains (whose diameter is smaller than the geometric length-scale $\frac{1}{b_0}$) scales as $E\propto M^2$ \cite{NE18}. We note that the recently derived  full compatibility conditions for three dimensional liquid crystals \cite{SE21,PA21} are associated with $\eta=1$ leading to $E\propto M^{5/3}$, which to the best of our knowledge has not been previously described or validated. 

\subsection{Bi-stable ribbon, the case of anisotropic domains and stiff Hamiltonians \label{bi-rib}}

The particularly well studied example of the frustrated bi-stable elastic ribbon \cite{GB05,AEKS11,AAMS14,GSD16,ZGDS19} allows three types of analysis: as a three dimensional elastic body, a thick two dimensional narrow ribbon, and a thin ribbon, highlighting the differences in the compatibility conditions for each of these limits. The system consists of a thin and narrow elastic ribbon, whose elastic energy may be expressed through its mid-surface metric, $a$, and second fundamental form $b$ through
\[
E=\int_{0}^{L}\int_{-w/2}^{w/2} \left( t |a-\bar{a}|^2+t^3 |b-\bar{b}|^2\right)dx^1dx^2,
\]
where $t,w$ and $L$ are the ribbon's thickness, width and length respectively. The reference values of the metric and second fundamental form read
\[
\bar{a}=\mymat{1&0\\0&1},\quad \bar{b}=\mymat{k_0&0\\0&-k_0},
\]
where $k_0^{-1}$ is the length-scale associated with the geometric frustration. 
For more details the reader is referred to  \cite{AEKS11,AAMS14}, and appendix \ref{snap}. 
\\
The compatibility conditions associated with the metric and second fundamental form consist of three distinct equations: one of the equations is algebraic in $b$ and involves second order derivatives of $a$ while the other two equations are first order differential equations relating the components of $a$ and $b$, see appendix \ref{snap}.

Fixing the thickness $t$ and considering the simultaneous variation of $w$ and $L$ keeping their ratio constant, in the regime $t<w<\sqrt{t/k_0}$, results in the thin and narrow limit of the frustrated ribbon, considered as a two dimensional surface. The leading order elastic energy in this case reads 
\[
E\propto k_0^4 t L w^5 \propto  t k_0^4 L^6\propto M^{3}, 
\]
in agreement with $\eta=2$ predicted by the compatibility conditions (see appendix \ref{snap}), similarly to the purely isotropic case considered in panel (a) of Figure (\ref{fig:elastic}). 
\\
We note that, while less intuitive, we may also consider this system as a three dimensional elastic structure, frustrated by the rest-length gradients across the ribbon's thickness. One may show that above energy indeed contains the leading order terms for the full three dimensional elastic energy, yet considered as a three dimensional object we must increase the systems thickness in proportion to its width and length yielding   
\[
E\propto k_0^4 t L w^5 \propto k_0^4 L^7\propto M^{7/3},
\]
in agreement with $\eta=2$ predicted by the three dimensional compatibility conditions (see appendix \ref{snap}), and again similarly to the purely isotropic case considered in panel (b) of Figure (\ref{fig:elastic}).
\\
We note that these two cases, while obtained from the same elastic energy and similarly behaving compatibility conditions, describe different physical settings. In the latter case the system grows isotropically in all three dimensions, while in the former case growth across the thin dimension is inhibited and only occurs laterally. 

The thin limit, in which in plane deformations are significantly more expensive energetically compared with bending deformations, violates the assumptions that the Hamiltonian contains no stiff directions. To study this limit within the present theory we are required to derive a new (reduced) Hamiltonian for the sought limit, in which there will no longer be any stiff directions. For thin elastic sheets the limiting elastic energy is well established, both by asymptotic expansions \cite{ESK09a} and by rigorous $\Gamma$-limits \cite{LP11}. The resulting limiting energy consists only of the bending energy, yet the solution space is now restricted only to isometries of the two dimensional metric. This manifests in the compatibility conditions in which the metric is now considered as a given quantity. As a result Gauss' equation becomes algebraic in the generalized strains and $\eta=0$, see appendix \ref{snap}.  
Thus, the thin limit in this case leads to an elastic energy that scales linearly with the total mass of the system.
\\
The limiting elastic energies for both 
thick and thin sheets are
associated with well defined  compatibility conditions and possess no stiff directions allowing us to predict $\lambda$ for these cases. However, describing the continuous transition between these two limits \cite{GB05,GSD16} remains outside the present framework. 

\section{\label{discussion}Discussion\protect\\ }

The framework provided in this work allows to quantify the strength of geometric frustration in physical assemblies. In particular, it provides a mean to distinguish cumulative from non-cumulative geometric frustration. 
Within this framework the compatibility conditions assume a central role in predicting the behavior of the system and in particular predict the super-extensive energy growth rate, $E\propto M^\lambda$. The compatibility conditions allow to identify the set of admissible states, which in turn greatly reduce the dimensionality of the problem. This dimensional reduction paves the way for better understanding frustrated assemblies, and may guide the engineering of their response \cite{GAE19,UE20}. We note that while the scaling of the generalized strain $\epsilon\propto r^\eta$ may be fractional (where $\eta$ is not an integer), in general, as the compatibility conditions related smooth fields we expect $\eta$ to be an integer. As a result the super extensive energy scaling exponent assumes but a few typical values:
$\lambda={1,5/3,7/3,3,...}$ for $3$-dimensional systems and $\lambda={1,2,3,4,...}$ for $2$-dimensional systems.

The scaling arguments that are used for determining the super-extensive energy growth rate presume that the domain considered is isotropic and that the associated Hamiltonian possesses no stiff directions.
To predict the behavior of systems with stiff directions (or domains of large aspect ratio), one should study the appropriate infinitely stiff limit (e.g. the vanishing thickness limit for the case of thin elastic sheets \cite{LP11,KS14}). The resulting dimensionally reduced Hamiltonian and compatibility conditions satisfy the framework's assumptions, and may be used to predict the system's behavior in this limit.
The energy scaling of the limiting Hamiltonian need not match that of the full Hamiltonian from which it was reduced. For example, as discussed in the previous section, while the full Hamiltonian of the non-Euclidean elastic bi-layer predicts a super extensive energy scaling, the reduced infinitely thin limit predicts an extensive energy scaling. 

The infinitely stiff (or diverging aspect ratio) limit may predict uniform frustration energy accumulation. However, in realistic physical systems, which are associated with a large but finite stiffness ratio (or aspect ratio), the geometric incompatibility may lead to spatial structural gradients in confined regions within the system, such as in the vicinity of its boundaries \cite{ESK09a,GBZ+12,KL17,MHR+20}.

The geometric length scale that arises from the compatibility conditions and locally preferred value of $\bar{\Psi}$ determine the domain size in which the strain follows a power-law growth, and restrict the predicted super extensive behavior to small domains. 
As the system's domain size approaches this length-scale the rate of strain accumulation may deviate from the predicted power law. In particular, the system's energy scaling may gradually become extensive. This is the case, for example, for large domains of twisted molecular crystals that gradually straighten as they grow \cite{HAS+19,SDS+20}, as well for bent core liquid crystals in the plane that approach the nematic texture for large domains.  
Note, however, that considering finite size assemblies, other mechanisms for locally relaxing the accumulated frustration may come into play. Such mechanisms include 
growth arrest in some of the spatial directions leading to filamentation \cite{LW17,HG17}, as well as the incorporation of point, line and surface defects that locally absorb the geometric charge. Such defects support discontinuities in the gradients of the fields describing the system allowing it to assume distinct solutions in the domains separated by the (possibly regularly spaced) defects \cite{AKNN05,MAK09,Gra12,LSS21,Sel21}.
The same system may display growth arrest along one direction, as well as introduce structural defects to attenuate the energy growth depending on the system's parameters \cite{MR20}.

Determining which of these  super-extensivity-saturation
mechanisms occur first requires a more delicate thermodynamic analysis of the system.  To predict, for example, if geometric frustration in a defect free structure will cause growth-inhibition and result in a self-limited open boundary structure, one should balance the rate of increase of the energy associated with the frustration induced distortions against the diminishing cost of surface energy per unit mass. Such a detailed thermodynamic analysis was recently carried out in \cite{HG20}.

The treatment of frustration presented here uses continuous fields to describe frustrated systems.
The frustration is captured by the compatibility conditions that prevent these fields from assuming their locally favored values, by enforcing local constraints between the fields and their spatial derivatives. Thus, the associated frustration, as well as the resulting compromise can be made arbitrarily small by considering infinitesimal domains. The discrete nature of the building blocks of realistic frustrated assemblies preclude such an arbitrarily small frustration and associated compromise. For example, the degrees of freedom in an Ising anti-ferromagnet on triangular lattice describing a variety of frustrated systems in condensed matter \cite{Wan50,KN53}, is associated with only two states. The inability of this system to distribute the desired compromise results in frustration saturation at the level of a single unit cell \cite{RL19}. Frustrated planar assemblies of elastic pentagons or heptagons allow continuous response, and exhibit cumulative frustrations \cite{LW17,thesisMeiriMSc2021}. However, the finite angle deficit associated with such non-tessellating polygons leads to frustration of large magnitude which in turn may result in deviations from the  predictions provided above for continuous systems. As these systems can be embedded in appropriately chosen uniformly curved spaces with no frustration one may apply the framework presented here to find a distortion minimizing embedding of finite sections of these curved spaces into $\mathbb{R}^3$. Possible candidates for energy minimizing conformations could then be obtained by applying the distortion minimizing map to the relaxed configuration in curved space. Such an approach was recently applied to study the frustrated assembly of  tetrahedral nano-particles \cite{SLK+21}. 
Some frustration mechanisms in discrete assemblies, such as the relative twist rate in fibril assemblies \cite{BKR14,HG17}, and the curvature in bent-core liquid crystals \cite{Sel21} can be continuously tuned to arbitrarily diminish the magnitude of the associated frustration. In such systems the continuous analysis presented here may accurately capture the system's behavior despite the discrete nature of the building blocks. The quantification of frustration provided by the present framework explains the origin of super extensive energy accumulation in continuous systems. Further analysis may relax the assumption of power law growth of the strain allowing treatment of highly frustrated discrete systems as well. Nonetheless the tension between the locally preferred arrangement and the long range order and the resulting  cooperative nature of the ground state of the system, are already well captured by the present continuous framework.

\begin{acknowledgments}
This work was funded by the Israel Science Foundation Grant No. 1479/16. EE thanks the Ascher Foundation for their support. 
\end{acknowledgments}

\appendix

\section{Slap bracelet} \label{snap}
We start by providing a three dimensional metric description to the elastic system. The two dimensional reductions will follow.
We parameterize the solid using its mid-surface $\mathbf{r}^{2D}(x^1,x^2)$ and the associated normal vector $\hat{\mathbf{N}}(x^1,x^2)$ through
\[
\mathbf{r}(x^1,x^2,x^3)=\mathbf{r}^{2D}(x^1,x^2)+x^3\hat{\mathbf{N}}(x^1,x^2) .
\]
The three dimensional metric in this case reads:
\[
g=\mymat{a_{11}&a_{12}&0\\
a_{12}&a_{22}&0\\
0&0&1}-2x^3 \mymat{b_{11}&b_{12}&0\\
b_{12}&b_{22}&0\\
0&0&1}+(x^3)^2\mymat{c_{11}&c_{12}&0\\
c_{12}&c_{22}&0\\
0&0&1},
\]
where $a_{\a\b}$, and $b_{\a\b}$ are the components of the metric and second fundamental form of the mid-surface, and the term quadratic in $x^3$ reads $c_{\a\b}=b_{\a\g}b_{\b\d}a^{\g\d}$. Note that $a,b$ and $c$ are all independent of $x^3$.

The intrinsic geometry of the bi-stable elastic ribbon also known as the snap bracelet (or slap bracelet) favors cylindrical conformations with one of two possible curvatures, equal in magnitude yet oriented towards opposite faces of the ribbon. These tendencies may be captured by setting 
\begin{equation}
\bar{g}=\mymat{1&0&0\\0&1&0\\0&0&1}
-2x^3\mymat{\kappa_0&0&0\\0&-\kappa_0&0\\0&0&0}.
\label{eq:gbarBilayer}
\end{equation}
One can show that eq. \ref{eq:gbarBilayer} captures the leading orders contributions for the reference metric that result from gluing two identical rectangular layers, each strained uniaxially before gluing along the longitudinal and transverse directions respectively \cite{AEKS11,CGM+12,KS14}.
A straightforward calculations yields that the Riemannian curvature tensor of $\bar{g}$ does not vanish. The three dimensional elastic energy 
\[
E=\frac{1}{2}\iiint A^{ijkl} \tfrac{1}{2}(g_{ij}-\bar{g}_{ij})\tfrac{1}{2}(g_{kl}-\bar{g}_{kl})d\bar{V},
\]
thus cannot be set to vanish on a finite domain. The corresponding compatibility conditions, embodied in the independent components of the Ricci curvature tensor, lead to a set of second order differential equations that predict $\eta=2$. This in turn leads to $\lambda=7/3$.

Within the same parametric regime (namely, $t<w<\sqrt{t /\kappa_0}$, and $w<1/\kappa_0$, \cite{AEKS11,AAMS14})
we may consider the system as a moderately thin sheet, which is described by its metric, $a$, and second fundamental form, $b$. In this regime 
\[
b\approx \bar{b}=\mymat{\kappa_0&0\\0&-\kappa_0},\quad \text{and} \quad
a \ne \bar{a}=\mymat{1&0\\0&1}.
\]
The compatibility conditions that relate the component of the metric and the second fundamental form are known as the Gauss-Peterson-Mainardi-Codazzi equations \cite{GAS97}:
\[
|b|=|a| K(a),\quad
\nabla_\a b_{\b\g}=\nabla_\b b_{\a\g},
\]
where $K(a)$ is the Gaussian curvature of the mid-plane as calculated from the Riemannian curvature tensor of the metric $a$. Gauss' equation relates the components of $b$ to $K(a)$ which includes second derivatives of the metric. The remaining two equations are first order in both $a$ and $b$. Na\"ively one may expect this to lead to $\eta=1$, yet expanding these equations in orders of the generalized strain $\Psi$ yields that the leading order equations are homogeneous resulting again in $\eta=2$. Other (and in particular, non-uniform) choices of $\bar{a}$ and $\bar{b}$ may lead to a different exponent $\eta$. As this system is considered two dimensional (at some constant thickness $t$) we obtain $\lambda =3$. 

In order to examine the limit of $t\to 0$ we are required to obtain the corresponding limiting energy functional, which minimizes the bending energy with respect to all isometric configurations \cite{ESK09,LP11,KS14}.   
%
The compatibility conditions in this case relate the components of the second fundamental form  to each other through:
\[
b_{11}b_{22}-b_{12}^2=0,\quad 
\partial_1 b_{12}=\partial_2 b_{11},\quad
\partial_1 b_{22}=\partial_2 b_{12}.
\]
In this case the non-homogeneous algebraic terms in Gauss' equation (proportional to $\kappa_0$) are not compensated by higher order terms and thus in this limit $\eta=0$. 

Indeed, solving for the the ground state of such a thin ribbon results in a surface that obeys the principle curvature directions of the reference curvatures, setting $b_{12}=0$. Energy minimization respecting Gauss' constraint reads $b_{11}=0$ and $b_{22}=-\bar{\kappa}_0$ or $b_{22}=0$ and $b_{11}=\bar{\kappa}_0$, which in turn result in an extensive bending energy (to leading order). For more details the reader is referred to \cite{AEKS11}, and its supplementary materials section.

\section{Embedding a geodesic disc  from $\mathbb{S}^{2}$ in $\mathbb{E}^{2}$} \label{app:Elas2D}
Consider a geodesic disc of radius $R$ cut from $\mathbb{S}^{2}$ parameterized through a polar semi-geosedic parameterization
\[
\bar{a}=
\begin{pmatrix}
1&0\\0&R^{2}\sin(r/R)^{2}
\end{pmatrix},
\]
where, for transparency we explicitly retain the curvature radius $R$ associated with the constant Riemannian curvature denoted by 
$K=R^{-2}$. Any planar embedding may be parametrized through the standard polar semi-geodesic coordinates which may be in turn recast using the reference coordinates through
\[
ds^{2}=d\rho^{2}+\rho^{2}d\theta^{2}=\rho'(r)^{2}dr^{2}+\rho(r)^{2}d\theta^{2}
\]
The covariant components of the strain thus read
\begin{equation}
\varepsilon=\frac{1}{2}\begin{pmatrix}
\rho'^{2}-1&0\\0&\rho^{2}-R^{2}\sin(r/R)^{2}
\end{pmatrix}.
\label{eq:strain}
\end{equation}
We next make use of the metric description of elasticity \cite{ESK09,ESK13}, which is particularly suited for residually stressed solids. 
For simplicity we consider an elastic media of vanishing Poisson ratio, and renormalize the Young's modulus to unity to obtain
\[
E=\int_{0}^{r_{max}}\int_{0}^{2\pi} \varepsilon^{\a}_{\b}\varepsilon^{\b}_{\a}R\sin[r/R] d\theta dr,
\]
where the mixed index strain tensor $\varepsilon_{\a}^{\b}=\bar{a}^{\a\g}\varepsilon_{\g\b}$ reads 
\[
\bar{a}^{-1}\varepsilon
=\frac{1}{2}\begin{pmatrix}
\rho'^{2}-1&0\\0&\frac{\rho^{2}}{R^2\sin(r/R)^{2}}-1
\end{pmatrix}.
\]
The elastic energy reduced to

\begin{align}
E= &\frac{\pi}{2}\int_{0}^{r_{max}} \left( 
(\rho'^{2}-1)^{2}+\left(\frac{\rho^{2}}{R^2\sin(\tfrac{r}{R})^{2}}-1\right)^{2}
\right)\cdot \notag \\
 & R\sin[\tfrac{r}{R}]  dr, \notag 
\end{align}
Numerically minimizing the above functional yields $E\propto r_{max}^{6}$, as can be observed in figure \ref{fig:elastic}.

We next come to consider the compatibility conditions that must be satisfied for a metric $a$ (and consequently the strain $\varepsilon=\tfrac{1}{2}(a-\bar{a})$) to describe a valid configuration in $\mathbb{R}^2$. The necessary and sufficient conditions in this case are the vanishing of all components in the Riemann curvature tensor. For two dimensions (2D) this yields just one equation, proportional to the Gaussian curvature of the metric $a$. 

The form of the strain that appears in equation \ref{eq:strain} and that is subsequently used in the numerical minimization comes from a configuration and thus is, by definition, compatible. To unveil the frustrated nature of this system we na\"ively write
$
a=\varepsilon +\bar{a},
$ making no assumptions regarding the structure of the strain, $\varepsilon$. We expand the compatibility condition expressed in terms of the strain in orders of the spatial coordinates. To zeroth order we obtain the non-homogeneous equation
\[
\Omega(\partial_\a \partial_\b \epsilon,\partial_\a \epsilon,\epsilon,\bar{a})+\frac{1}{R^2}=0,
\]
where the highest order of strain derivatives that the functions $\Omega$ depends on, are second derivatives leading to $\eta=2$.

We now repeat the above exercise with a different measure of strain and a different elastic energy. As the system displays small strains, we expect all descriptions to agree and in particular expect to obtain a similar scaling exponent. The elastic energy we employ follows from a deformation gradient approach, which yields an explicitly solvable Euler Lagrange equation. In this approach the basic variable is the deformation gradient $F=\nabla \mathbf{r}$, which satisfies $F^{T}F=g$. The deformation gradient is compared with a reference value $\bar{F}$ that satisfies $\bar{F}^{T}\bar{F}=\bar{g}$, and is defined up to a rigid rotation. This reference value is termed the ``virtual'' deformation gradient \cite{GN71,Hog97}, as in frustrated systems it does not correspond to a gradient of a configuration.
Rigid motions, and in particular rigid rotations, lead to no elastic distortions yet vary $F$. This manifests in the elastic energy, which measures the distance of $\bar{F}^{-1}F$ from $SO(2)$, eliminating the associated freedom of a rigid rotation from the elastic energy:
\begin{equation*}
E=\int \text{dist}^{2}\left(\bar{F}^{-1}F,SO(2)\right) d\bar{A},
\label{eq:EofF}
\end{equation*}
where $SO(2)$ denotes the family of orientation preserving rotations. This energy is better behaved mathematically compared with the metric description (as it penalizes local inversions), and thus is often favored as the starting point for formal $\Gamma-$limits calculations \cite{FJM06,KS14}. However, it is less accessible geometrically and often intractable analytically rendering it difficult to apply in many practical settings. In the present case, the high symmetry of the problem and expected solution implies that the radial and azimuthal directions keep their orientations. As a result we identify the identity as the member of $SO(2)$ closest to $\bar{F}^{-1}F$and obtain the simple energy
\[
E=\frac{\pi}{2} \int_{0}^{r_{max}} \left( 
(\rho'-1)^{2}+(\frac{\rho}{R\sin(\tfrac{r}{R})}-1)^{2}
\right)
R\sin[\tfrac{r}{R}]  dr.
\]
This leads to a linear Euler Lagrange equation  
\[
(\frac{\rho}{R\sin(\tfrac{r}{R})}-1)-\frac{d}{dr}\left((\rho'-1)R\sin[\tfrac{r}{R}]\right)=0,
\]
supplemented by the boundary conditions $\rho(0)=0$ and $\rho'(r_{max})=1$; the solution for which reads:

\begin{align}
\rho\left( r \right)  =&  -2 R \tan \left( \frac{u}{2 R} \right) \left( \cot^2 \left( \frac{r_\text{max}}{2 R} \right) \log \left( \cos \left( \frac{r_\text{max}}{2 R} \right) \right) \right.  \notag \\
 &\left. +\cot ^2\left(\frac{u}{2 R} \right) \log \left( \cos \left( \frac{u}{2 R} \right) \right) \right) . \notag 
\end{align}

Figure \ref{fig:elastic} shows that the elastic energy \ref{eq:EofF}
of this configuration leads to a similar energy growth exponent. 

The compatibility condition for $F$, however, lead to a first order set of equations, as they arise from $\nabla\times F=0$. To obtain the exponent $\eta$ from this approach we write the deformation gradient in terms of the
elastic strain $F=\varepsilon+\bar{F}$ (note that this $\varepsilon$ is not symmetric).  
We then write the compatibility conditions without any assumptions as to the form of $\varepsilon$. For transparency we will use a Cartesian curl operator, which requires that we express $\bar{F}$ in its Cartesian form 

\begin{align}
&\bar{F}_{cart}= O^T\bar{F}J=   \notag \\
 &\kappa^{-3/2}\mymat{
x^2\sqrt{\kappa}+R y^2 \sin\left( \frac{\sqrt{\kappa}}{R}\right)&
xy\sqrt{\kappa}-R xy \sin\left(\frac{\sqrt{\kappa}}{R}\right)\\
xy\sqrt{\kappa}-R xy \sin\left(\frac{\sqrt{\kappa}}{R}\right)&
y^2\sqrt{\kappa}+R x^2 \sin\left(\frac{\sqrt{\kappa}}{R}\right)},  \notag 
\end{align}
where $O=\mymat{\cos(\theta)&\sin(\theta)\\ -\sin(\theta)& \cos(\theta)}$ is the rotation from Cartesian to polar directions, $J=\frac{\partial(r,\theta)}{\partial(x,y)}$ is the Jacobian matrix associated with the coordinate transformation and $\kappa=x^2+y^2$ . The equations $\nabla\times F=0$ read 
\begin{align}
 0&=
\mymat{\partial_y \varepsilon_{11}-\partial_x \varepsilon_{12}\\
\partial_y \varepsilon_{21}-\partial_x \varepsilon_{22}}+x\mymat{\partial_x\partial_y \varepsilon_{11}-\partial_x\partial_x \varepsilon_{12}\\
\frac{1}{2t^2}+\partial_x\partial_y \varepsilon_{21}-\partial_x\partial_x \varepsilon_{22}}  \notag \\
&+y\mymat{-\frac{1}{2t^2}+\partial_y\partial_y \varepsilon_{11}-\partial_x\partial_y \varepsilon_{12}\\
\partial_y\partial_y \varepsilon_{21}-\partial_x\partial_y \varepsilon_{22}}+\mathcal{O}(x^2+y^2),   \notag
\end{align}

where all the derivatives of the strain components are estimated at the origin, $x=y=0$. 
Note that in this case the zero order of the compatibility condition is homogeneous, making the first order (which includes second derivatives of the strain) the leading non-homogeneous term. We may thus claim that $\eta=2$ in this case as well. 

\section{Embedding a geodesic disc  from $\mathbb{S}^{3}$ in $\mathbb{E}^{3}$} \label{app:Elas3D}
Similarly to the procedure carried out for the 2D case we begin by considering a finite domain cut from $\mathbb{S}^{3}$ parameterized by spherical coordinates. With respect to these coordinates the reference metric reads 
\[
\bar{g}=
\begin{pmatrix}
1&0&0\\0& R^{2}\sin(r/R)^{2} \sin(\phi)^2&0\\0&0&R^{2}\sin(r/R)^{2}
\end{pmatrix}.
\]
Again, we assume that the embedding preserves the spherical symmetry and thus can be given in terms of a single radial function $\rho$

\begin{align}
ds^{2}=&d\rho^{2}+\rho^{2}\sin(\phi)^2d\theta^{2}+
\rho^{2}d\phi^{2}=   \notag \\
 &\rho'(r)^{2} dr^{2}+\rho(r)^{2}\sin(\phi)^2d\theta^{2}+\rho(r)^{2}d\phi^{2}.  \notag 
\end{align}
The resulting elastic energy thus reads
\begin{align}
E=&\pi \int_{0}^{r_{max}} \left( 
(\rho'^2-1)^{2}+2(\frac{\rho^2}{R^2\sin(\tfrac{r}{R})^2}-1)^{2}
\right) \cdot  \notag \\
 &R^2\sin[\tfrac{r}{R}]^2  dr.  \notag 
\end{align}
The minimal values of the above energy for various values of $r_{max}$ are presented in Figure \ref{fig:elastic}, and follow $E\propto M^{7/3}$, where $M$ (reference volume) is the mass of the region considered. 

Similarly to the calculation carried out for the 2D case, in order to obtain the form of the compatibility conditions one needs to consider a na\"ive approach where the form of metric of the configuration $g$ is not presumed to come from an embedding. The compatibility conditions again correspond to the vanishing of all components of the Riemann curvature tensor, yet for three dimensional (3D) manifolds there are six independent components (from which we can construct three independent scalar equations). The resulting relations, much like the case for 2D, are second order differential equations whose zeroth order (expanded in the spatial coordinates) yields a non-homogeneous relation. Thus, here as well we obtain $\eta=2$ and the exponent $7/3$ follows. 

Last, we come to consider the same problem approached using a deformation gradient formulation. The energy reads

\[
E=\pi \int_{0}^{r_{max}} \left( 
(\rho'-1)^{2}+2(\frac{\rho}{R\sin(\tfrac{r}{R})}-1)^{2}
\right)
R^2\sin[\tfrac{r}{R}]^2  dr.
\]
Figure \ref{fig:elastic} presents the minimal values of the above energy for various $r_{max}$ values, following the same exponent as the metric description. 

Much like the 2D case, the compatibility conditions here form a linear set of equations as they too arise from $\nabla\times F=0$. However, for 3D these result in nine equations. We again seek to implement the curl in Cartesian coordinates. We thus write $\bar{F}_{cart}=O^T\bar{F}J$, where
\begin{align}
 O=&\mymat{\cos(\theta)\sin(\phi)& \sin(\theta)\sin(\phi)& \cos(\phi)\\
-\sin(\theta)&\cos(\theta)&0\\
\cos(\theta)\cos(\phi)&\sin(\theta)\cos(\phi)&-\sin(\phi)}\qquad \text{and} \notag\\  
 J=&\frac{\partial (r,\theta,\phi)}{\partial (x,y,z)}, \notag   
\end{align}
are the rotation matrix transforming between Cartesian and spherical directions, and the associated Jacobian matrix, respectively. The zeroth order of the nine equations that arise from $\nabla\times (\bar{F}_{cart}+\varepsilon)=0$ yields only homogeneous first order differential equation for the strain, much like the case for 2D. The next order (linear in the coordinates) yields non-homogeneous equations and thus gives as well $\eta=2$.

\bibliographystyle{unsrt}
\bibliography{main.bib}

\end{document}